\documentclass{epl}

\newcommand{\nc}{\newcommand}
\nc{\beq}{\begin{equation}}
\nc{\eeq}{\end{equation}}
\nc{\beqa}{\begin{eqnarray}}
\nc{\eeqa}{\end{eqnarray}}
\nc{\lra}{\leftrightarrow}

\nc{\sss}{\scriptscriptstyle}
{\nc{\lsim}{\mbox{\raisebox{-.6ex}{~$\stackrel{<}{\sim}$~}}}
{\nc{\gsim}{\mbox{\raisebox{-.6ex}{~$\stackrel{>}{\sim}$~}}}

\title{A Comment on ``Brans-Dicke Cosmology with a scalar field potential"}
\author{J\'er\'emie Vinet\inst{1}}
\institute{
  \inst{1} Physics Department, McGill University, Montr\'eal, Qu\'ebec, Canada H3A 2T8\\
  }

\begin{document}

\maketitle

\begin{abstract}
We show that a recent letter claiming to present exact cosmological solutions in Brans-Dicke theory actually uses a flawed set of equations as the starting point for their analysis.  The results presented in the letter are therefore not valid.
\end{abstract}

\section{Introduction}
In a recent letter, Mak and Harko \cite{mak} presented a set of solutions to Brans-Dicke theory with a quartic potential and a perfect fluid matter distribution.  However, the equations which were the starting point of the analysis contain an error, which invalidates all subsequent results.

\section{Inconsistency in equations of motion}
A quick check of the literature (for just a few examples, see \cite{others}) dealing with cosmology in Brans-Dicke theories suffices to conclude that equation $(4)$ of \cite{mak} has the wrong sign in front of the term involving the potential and its derivative.  However, in the interest of completeness, we will show explicitly that using the Friedmann equations and conservation of energy $(\dot \rho = -3H(\rho +p))$ leads to an equation different from $(4)$ of \cite{mak}.  Furthermore, this is not the result of a typographical error, since as we will show, the starting point for the analysis, equation $(5)$ of \cite{mak}, depends on the wrong sign in $(4)$ of \cite{mak}.

We begin by combining equations $(2)$ and $(3)$ of \cite{mak} to write 
\beqa
3H^2 &=& \frac{\rho_m}{\phi}+\frac{\omega}{2}\psi^2+\frac{V(\phi)}{2\phi}-3H\psi
\eeqa
and
\beqa
2\dot H +3H^2 &=& -\frac{p_m}{\phi}-\left(1+\frac{\omega}{2}\right)\psi^2+\frac{V(\phi)}{2\phi}-\dot\psi-2H\psi\\
\Rightarrow \dot H &=&\frac{1}{2}\left[-\frac{\rho_m+p_m}{\phi}-\left(1+\omega\right)\psi^2-\dot\psi+H\psi\right]
\eeqa
where $\psi \equiv \dot\phi/\phi$.  Taking the time derivative of the first of these leads to
\beqa
6H\dot H &=& \frac{\dot \rho_m}{\phi}-\frac{\rho_m}{\phi}\psi+\omega\psi\dot\psi+\frac{V'(\phi)}{2}\psi-\frac{V(\phi)}{2\phi}\psi-3\dot H\psi-3H\dot\psi\\
3(2H+\psi)\dot H &=&-3\frac{H}{\phi}(\rho_m+p_m)-\psi\left(\frac{\rho_m}{\phi}-\omega\dot\psi-\frac{V'(\phi)}{2}+\frac{V(\phi)}{2\phi}\right)-3H\dot\psi
\eeqa
\beqa
\frac{3}{2}(2H+\psi) \left[-\frac{\rho_m+p_m}{\phi}-\left(1+\omega\right)\psi^2-\dot\psi+H\psi\right]&=&-3\frac{H}{\phi}(\rho_m+p_m)-3H\dot\psi\nonumber\\
&&-\psi\left(\frac{\rho_m}{\phi}-\omega\dot\psi-\frac{V'(\phi)}{2}+\frac{V(\phi)}{2\phi}\right) 
\eeqa 
where dots denote time derivatives, and primes denote derivative w.r.t. the scalar $\phi$. Collecting terms, we obtain
\beqa
-\frac{(\rho_m+3p_m)}{\phi}&=&(3+2\omega)\dot\psi+V'(\phi)-\frac{V(\phi)}{\phi}+3H\psi(1+2\omega)+3\psi^2(1+\omega)-6H^2\\
-\frac{(\rho_m+3p_m)}{\phi}&=&(3+2\omega)\dot\psi+V'(\phi)-\frac{V(\phi)}{\phi}+3H\psi(1+2\omega)+3\psi^2(1+\omega)\nonumber\\
&&-2\left(\frac{\rho_m}{\phi}+\frac{\omega}{2}\psi^2+\frac{V(\phi)}{2\phi}-3H\psi\right)
\eeqa  
which finally leads to the following equation
\beqa
\frac{(\rho_m-3p_m)}{\phi}+\left(2\frac{V(\phi)}{\phi}-V'(\phi)\right)=(3+2\omega)\left(\dot\psi+3H\psi+\psi^2\right).
\eeqa
We now see that the sign in front of the term containing the potential and its derivative is the opposite of that given in $(4)$ of \cite{mak}.  Let us now show that equation $(5)$ of \cite{mak}, which constitutes the starting point for their analysis, does not follow from the correct equation we have written here, but from their erroneous equation $(4)$ of \cite{mak}.

We start with the equation we have just written down, and use the Friedmann equations to replace $\rho_m$ and $p_m$
\beqa
(2\omega+3)(\dot\psi +\psi^2+3H\psi)&=&\left(3H^2-\frac{\rho_{\phi}}{\phi}\right)+\left(6\dot H +9H^2+3p_{\phi}\right)+2\frac{V(\phi)}{\phi}-V'(\phi)\\
(2\omega+3)(\dot\psi +\psi^2+3H\psi)&=&12 H^2 +6\dot H+2\frac{V(\phi)}{\phi}-V'(\phi)-\left(\frac{\omega}{2}\psi^2+\frac{V(\phi)}{2\phi}-3H\psi\right)\nonumber\\
&&+3\left(\left(1+\frac{\omega}{2}\right)\psi^2-\frac{V(\phi)}{2\phi}+\dot\psi+2H\psi\right),
\eeqa
which, once we collect all terms, leads to
\beqa
\dot\psi-\frac{3}{\omega}\dot H = \frac{6}{\omega}H^2-\frac{1}{2}\psi^2-3H\psi-\frac{V'(\phi)}{2\omega}.
\eeqa
Thus the claim made in \cite{mak} that choosing $V(\phi) = V_0\phi^4$ allows the contributions of the potential and its derivative to cancel in the above equation is clearly wrong since only the derivative appears in the correct equation.  Since all subsequent results depend on this faulty affirmation, they are not valid.

\end{document}